
\input harvmac

\rightline{RI-1-00}
\rightline{CERN-TH/2000-052}
\Title{
\rightline{hep-th/0002104}
}
{\vbox{\centerline{Superstring Theory on $AdS_3 \times G/H$ and}
\centerline{Boundary $N=3$ Superconformal Symmetry}}}
\medskip
\centerline{\it Riccardo Argurio$^1$, Amit Giveon$^{1,2}$ and 
Assaf Shomer$^1$}
\vskip .3in
\centerline{{}$^1$ Racah Institute of Physics}
\centerline{The Hebrew University}
\centerline{Jerusalem 91904, Israel}
\vskip .2in
\centerline{{}$^2$ Theory Division, CERN}
\centerline{CH-1211, Geneva 23, Switzerland}
\vskip .2in

\centerline{\tt argurio, shomer@cc.huji.ac.il, giveon@vms.huji.ac.il}

\vskip .5in

\noindent
Superstrings propagating on backgrounds of the form 
$AdS_3 \times G/H$ are studied using the coset CFT approach. 
We focus on seven dimensional
cosets which have a semiclassical limit, and which give rise to
$N=3$ superconformal symmetry in the dual CFT. 
This is realized for the two cases
$AdS_3 \times SU(3)/U(1)$ and $AdS_3 \times SO(5)/SO(3)$, for which we
present an explicit construction. 
We also provide sufficient conditions on a CFT background to
enable a similar construction, and comment on the geometrical
interpretation of our results.

\Date{2/00}

\def\journal#1&#2(#3){\unskip, \sl #1\ \bf #2 \rm(19#3) }
\def\andjournal#1&#2(#3){\sl #1~\bf #2 \rm (19#3) }

\def\frac#1#2{{#1\over#2}}

\def\half{\frac12}

\def\inbar{\,\vrule height1.5ex width.4pt depth0pt}
\def\IC{\relax\hbox{$\inbar\kern-.3em{\rm C}$}}
\def\IR{\relax{\rm I\kern-.18em R}}
\def\IP{\relax{\rm I\kern-.18em P}}

%
%

%
\catcode`\@=11
\def\slash#1{\mathord{\mathpalette\c@ncel{#1}}}
\overfullrule=0pt

\def\NN{{\cal N}}

\def\eps{\epsilon}

\def\underrel#1\over#2{\mathrel{\mathop{\kern\z@#1}\limits_{#2}}}

\catcode`\@=12


%


\newsec{Introduction}
String propagation on curved backgrounds with an $AdS_3$ factor
has been of recent  interest %
\nref\ms{J.~Maldacena and A.~Strominger,
``$AdS_3$ black holes and a stringy exclusion principle,''
JHEP {\bf 9812} (1998) 005,
hep-th/9804085.}%
\nref\egp{J.~M.~Evans, M.~R.~Gaberdiel and M.~J.~Perry,
``The no-ghost theorem for $AdS_3$ and the stringy exclusion principle,''
Nucl.\ Phys.\  {\bf B535} (1998) 152,
hep-th/9806024.}%
\nref\gks{A.~Giveon, D.~Kutasov and N.~Seiberg,
``Comments on string theory on $AdS_3$,''
Adv.\ Theor.\ Math.\ Phys.\  {\bf 2} (1998) 733,
hep-th/9806194.}%
\nref\dort{J.~de Boer, H.~Ooguri, H.~Robins and J.~Tannenhauser,
``String theory on $AdS_3$,''
JHEP {\bf 9812} (1998) 026,
hep-th/9812046.}%
\nref\kuse{D.~Kutasov and N.~Seiberg,
``More comments on string theory on $AdS_3$,''
JHEP {\bf 9904} (1999) 008,
hep-th/9903219.}%
\nref\mo{J.~Maldacena and H.~Ooguri,
``Strings in $AdS_3$ and SL(2,R) WZW model: I,''
hep-th/0001053.}%
\nref\oldrefs{
A. B. Zamolodchikov and V. A. Fateev, Sov. J. Nucl. Phys. {\bf 43}
(1986) 657;
J. Balog, L. O'Raifeartaigh, P. Forgacs, and A. Wipf,
Nucl. Phys. {\bf B325} (1989) 225;
L. J. Dixon, M. E. Peskin and J. Lykken,  Nucl. Phys. {\bf B325} (1989)
329;
A. Alekseev and S. Shatashvili, Nucl. Phys. {\bf B323} (1989) 719;
N. Mohameddi, Int. J. Mod. Phys. {\bf A5} (1990) 3201;
P. M. S. Petropoulos, Phys. Lett. {\bf B236} (1990) 151;
M. Henningson and S. Hwang, Phys. Lett. {\bf B258} (1991) 341;
M. Henningson, S. Hwang, P. Roberts, and B. Sundborg, Phys. Lett. {\bf
B267} (1991) 350;
S. Hwang, Phys. Lett. {\bf B276} (1992) 451, hep-th/9110039;
I. Bars and D. Nemeschansky, Nucl. Phys. {\bf B348} (1991) 89;
S. Hwang, Nucl. Phys. {\bf B354} (1991) 100;
K. Gawedzki, hep-th/9110076;
I. Bars, Phys. Rev. {\bf D53} (1996) 3308, hep-th/9503205;
in {\it Future Perspectives In String Theory} (Los Angeles, 1995),
hep-th/9511187;
O. Andreev, hep-th/9601026, Phys.Lett. {\bf B375} (1996) 60,
Nucl.\ Phys.\  {\bf B552} (1999) 169, hep-th/9901118,
Nucl.\ Phys.\  {\bf B561} (1999) 413, hep-th/9905002, hep-th/9909222;
J. L. Petersen, J. Rasmussen and M. Yu, hep-th/9607129, Nucl.Phys. {\bf
B481} (1996) 577;
Y. Satoh, Nucl. Phys. {\bf B513} (1998) 213, hep-th/9705208;
J. Teschner, hep-th/9712256,  hep-th/9712258, hep-th/9906215;
I.~Pesando, JHEP {\bf 9902} (1999) 007, hep-th/9809145, 
Mod.\ Phys.\ Lett.\  {\bf A14} (1999) 2561, hep-th/9903086;
J.~Rahmfeld and A.~Rajaraman, Phys.\ Rev.\  {\bf D60} (1999) 064014
hep-th/9809164;
S.~Mukherji and S.~Panda, Phys.\ Lett.\  {\bf B451} (1999) 53, hep-th/9810140;
K.~Ito, Phys.\ Lett.\  {\bf B449} (1999) 48, hep-th/9811002,
Mod.\ Phys.\ Lett.\  {\bf A14} (1999) 2379, hep-th/9910047;
K.~Hosomichi and Y.~Sugawara, JHEP {\bf 9901} (1999) 013, hep-th/9812100,
JHEP {\bf 9907} (1999) 027, hep-th/9905004;
M.~Yu and B.~Zhang, Nucl.\ Phys.\  {\bf B551} (1999) 425, [hep-th/9812216;
N.~Berkovits, C.~Vafa and E.~Witten, JHEP {\bf 9903} (1999) 018,
hep-th/9902098;
J.~de Boer, A.~Pasquinucci and K.~Skenderis, hep-th/9904073;
M.~Banados and A.~Ritz, Phys.\ Rev.\  {\bf D60}, 126004 (1999), hep-th/9906191;
I.~Bars, C.~Deliduman and D.~Minic, hep-th/9907087;
P.~M.~Petropoulos, hep-th/9908189;
Y.~Sugawara, hep-th/9909146;
G.~Giribet and C.~Nunez, JHEP {\bf 9911} (1999) 031, hep-th/9909149;
A.~Kato and Y.~Satoh, hep-th/0001063;
M.~Langham, hep-th/0002032.}%
(see for instance \refs{\ms-\mo}, and \oldrefs\ for additional references).
One motivation is the fact that $AdS_3\simeq SL(2)$ 
is an exact background
which can be treated in string pertubation theory, and thus
allows to consider the $AdS$/CFT correspondence \ref\malda{J.~Maldacena,
``The large N limit of superconformal field theories and supergravity,''
Adv.\ Theor.\ Math.\ Phys.\  {\bf 2} (1998) 231,
hep-th/9711200.}
beyond the supergravity limit. 
Some specific examples that were studied in this context include
superstrings propagating on $AdS_3 \times \NN$ where $\NN$ was
a group manifold %
\nref\efgt{S.~Elitzur, O.~Feinerman, A.~Giveon and D.~Tsabar,
``String theory on $AdS_3\times S^3 \times S^3 \times S^1$,''
Phys.\ Lett.\  {\bf B449} (1999) 180,
hep-th/9811245.}%
\refs{\gks,\efgt}, or an orbifold of a group manifold %
\nref\kll{D.~Kutasov, F.~Larsen and R.~G.~Leigh,
``String theory in magnetic monopole backgrounds,''
Nucl.\ Phys.\  {\bf B550} (1999) 183,
hep-th/9812027.}%
\nref\yis{S.~Yamaguchi, Y.~Ishimoto and K.~Sugiyama,
``$AdS_3$/CFT$_2$ correspondence and space-time N = 3 superconformal  
algebra,'' JHEP {\bf 9902} (1999) 026,
hep-th/9902079.}%
\refs{\kll,\yis}.
In this paper we study cases in which $\NN$ is a coset
manifold. This is an interesting generalization of the $AdS$/CFT correspondence
which has been considered in the higher dimensional cases of type IIB
string theory on $AdS_5 \times \NN^5$ 
\ref\kw{I.~R.~Klebanov and E.~Witten,
``Superconformal field theory on threebranes at a Calabi-Yau  singularity,''
Nucl.\ Phys.\  {\bf B536} (1998) 199,
hep-th/9807080.} 
and of M-theory on $AdS_4 \times \NN^7$ 
\nref\cast{L.~Castellani, A.~Ceresole, R.~D'Auria, 
S.~Ferrara, P.~Fre and M.~Trigiante,
``G/H M-branes and $AdS_{p+2}$ geometries,''
Nucl.\ Phys.\  {\bf B527} (1998) 142,
hep-th/9803039.}%
\nref\figu{B.~S.~Acharya, J.~M.~Figueroa-O'Farrill, C.~M.~Hull and B.~Spence,
``Branes at conical singularities and holography,''
Adv.\ Theor.\ Math.\ Phys.\  {\bf 2} (1999) 1249,
hep-th/9808014; J.~M.~Figueroa-O'Farrill,
``Near-horizon geometries of supersymmetric branes,''
hep-th/9807149; ``On the supersymmetries of anti de Sitter vacua,''
Class.\ Quant.\ Grav.\  {\bf 16} (1999) 2043,
hep-th/9902066.}%
\refs{\cast,\figu}, where $\NN^5$ and $\NN^7$ are Einstein manifolds
(generically coset manifolds)
preserving a fraction of supersymmetry.
This type of construction allows one to consider dual supersymmetric CFTs
which are not `orbifolds' of the maximally supersymmetric one.
The $AdS_3 \times \NN$ case is somewhat different since here we have the
possibility of studying $\NN$ in the context of coset CFTs.

We choose to study coset CFTs which have a large radius
(or large level $k$) semiclassical limit, 
corresponding to superstrings propagating on seven dimensional coset manifolds.
Moreover, we focus on cases in which the dual two dimensional
theory (also referred to as the spacetime CFT)
has an extended superconformal symmetry.\foot{In the following
we refer to the supersymmetry of, say, the left-movers only. The supersymmetry
of the other sector depends on the particular superstring theory considered.} 
Coset models leading
to $N=2$ can be easily realized 
as particular cases of the general construction of %
\nref\gr{A.~Giveon and M.~Ro\v cek,
``Supersymmetric string vacua on $AdS_3\times \NN$,''
JHEP {\bf 9904} (1999) 019,
hep-th/9904024;
D.~Berenstein and R.~G.~Leigh,
``Spacetime supersymmetry in $AdS_3$ backgrounds,''
Phys.\ Lett.\  {\bf B458} (1999) 297,
hep-th/9904040.}%
\refs{\gr}, where $\NN$ decomposes
as a $U(1)$ factor times a Kazama-Suzuki model 
\ref\ks{Y.~Kazama and H.~Suzuki,
``New N=2 Superconformal Field Theories And Superstring Compactification,''
Nucl.\ Phys.\  {\bf B321} (1989) 232; ``Characterization Of N=2 
Superconformal Models Generated By Coset Space Method,''
Phys.\ Lett.\  {\bf B216} (1989) 112.}. 
On the other hand, there are no seven dimensional
coset manifolds leading to $N=4$ supersymmetry in spacetime (except of
course the cases \refs{\gks,\efgt}
in which the cosets are actually group manifolds).
Therefore, we shall be interested in the cases where the spacetime
CFT has $N=3$ supersymmetry.

Our main result, presented in sections 3 and 4,  
is the construction of the spacetime $N=3$ superconformal
algebra in the two cases:
\eqn\adsu{AdS_3 \times {SU(3) \over U(1) }, \qquad \qquad
AdS_3 \times {SO(5) \over SO(3) },}
which actually turn out to be the only coset models giving rise to $N=3$.
An interesting by-product of the construction is related to the fact 
that getting $N=3$ 
depends on the choice of chiral GSO projection. 
Choosing the opposite projection
leads to an $N=1$ superconformal algebra in spacetime together with
an $SU(2)$ affine algebra acting trivially on the supercharges.
This spacetime structure also appears in \yis, where a $Z_2$ orbifold
of the large $N=4$ algebra obtained by \efgt\ is taken.

We also go beyond the coset set-up by providing in section 6
a set of sufficient conditions on a CFT background $\NN$ which allow for the
construction of $N=3$ superconformal
symmetry in spacetime from superstrings on $AdS_3 \times \NN$.
We recover the $N=1$ structure for the other GSO projection also in this
general set-up.
The proof elaborates on the construction of $N=2$ spacetime supersymmetry
by \gr, the additional ingredient being the presence in the 
CFT on $\NN$ of an affine $SU(2)$.
Finally, we comment in section 7
on a possible geometrical interpretation of these
conditions, and relate our work to the case of M-theory compactified on
$AdS_4 \times \NN^7$ \refs{\cast,\figu}, as well as to brane
configurations.

\newsec{Spacetime $N=3$ superconformal algebra}

Extended superconformal algebras in two dimensions
also include an affine R-symmetry algebra, which generally leads to
a quantization of the central charge in unitary theories.
Specifically, the $N=3$ superconformal algebra has an affine
$SU(2)$ subalgebra.
The central charge is related to the level $\tilde{k}$ of this affine $SU(2)$,
which is an integer, by $\tilde{c}={3\over 2} \tilde{k}$
\ref\ss{A.~Schwimmer and N.~Seiberg,
``Comments On The N=2, N=3, N=4 Superconformal Algebras In Two-Dimensions,''
Phys.\ Lett.\  {\bf B184} (1987) 191.}.
Therefore, a necessary condition for string theory on a background of
the form $AdS_3 \times \NN$ to have spacetime $N=3$ superconformal
symmetry is the existence of an affine $SU(2)$ in spacetime.
This is obtained when the worldsheet CFT on $\NN$ has an affine 
$SU(2)$ symmetry as well \gks. If the respective worldsheet
levels of $SL(2)$ and of $SU(2)$ are $k$ and $k'$, the analysis of
\refs{\gks,\kuse}
shows that in the spacetime theory we have $\tilde{c}=6kp$
and $\tilde{k}=k'p$, where $p$ is the integer number introduced %
\nref\sw{N.~Seiberg and E.~Witten, ``The D1/D5 system and singular CFT,''
JHEP {\bf 9904} (1999) 017, hep-th/9903224.}%
in \gks, related to the maximal number of `long strings' \refs{\kuse,\sw}.
A further condition is thus $k'=4k$ (recall that $k$ is not forced
to be an integer).

In the following we focus on coset manifolds $\NN$ which have 7 dimensions,
so that a large $k$ semi-classical limit is possible.
Two cases which satisfy the
conditions given above are $SL(2)_k \times \NN$ with:
\eqn\su{\NN_1 = {SU(3)_{4k} \over U(1) },}
and
\eqn\so{\NN_2 = {SO(5)_{4k} \over SO(3) }.}
Note that there are several ways of choosing the $SO(3)$ in $\NN_2$,
according to the nesting of subgroups 
$SO(3) \subset SO(3) \times SO(3) \simeq SO(4) \subset SO(5)$. 
Since we require $\NN_2$ to have an unbroken $SO(3)$
symmetry, we are forced to mod out by one of the two $SO(3)$ factors of
$SO(4)$.\foot{Note that modding out by the diagonal $SO(3)$ and
then by a further $U(1)$ would lead to a Kazama-Suzuki model \ks.}
It is straightforward to show that the two models above are critical:
\eqn\csu{c_{sl}+c_1=\left({9\over 2}
 +{6\over k}\right) + \left( 12 -{24\over 4k} -{3\over2}
\right) = 15,}
\eqn\cso{c_{sl}+c_2=\left({9\over 2} +{6\over k}\right)
 + \left( 15 -{30\over 4k}
-{9\over 2} + {6\over 4k} \right) =15.}
We now show that these two models indeed possess $N=3$ superconformal
symmetry in spacetime by explicit construction.
Since the construction is similar in the two cases, we will focus here
on the first case, and then go briefly over the second one.

\newsec{Superstring theory on $AdS_3 \times SU(3)/U(1)$}
We first have to set some notations, starting from the $SL(2)$ WZW part.
We mainly follow the formalism of \ks\ and \gks. For simplicity we
only treat the holomorphic sector.

The $SL(2)$ supersymmetric WZW model is constituted of the three currents
of the $SL(2)$ affine algebra at level $k$, and the three fermions
implied by the $N=1$ worldsheet supersymmetry, satisfying the following
OPEs:
\eqn\slope{\eqalign{J^P(z) J^Q(w) \sim&
{{k\over 2} \eta^{PQ} \over (z-w)^2} + {i\eps^{PQR}
\eta_{RS} J^S(w) \over z-w},\cr
J^P(z) \psi^Q(w) \sim &{i\eps^{PQR} \eta_{RS} \psi^S(w)
\over z-w},\cr
\psi^P(z) \psi^Q(w) \sim &{ {k\over 2}\eta^{PQ} \over z-w},}}
where $P,Q,R,S=1,2,3$, $\eta^{PQ}=(++-)$ and $\eps^{123}=1$.
As usual in supersymmetric WZW models, the currents can be decomposed in
two pieces:
\eqn\decomp{J^P=\hat{J}^P-{i\over k}\eta^{PQ}\eps_{QRS}\psi^R \psi^S.}
The first piece $\hat{J}^P$ constitutes an affine algebra at level $k+2$,
and has regular OPE with the fermions $\psi^P$. We will thus refer
to $\hat{J}^P$ as the bosonic currents. The second part constitutes
an affine algebra at level $-2$, and is referred to as the fermionic
part of the current.

The worldsheet stress-energy tensor and $N=1$ supercurrent are:
\eqn\sltg{\eqalign{T_{sl}=&{1\over k}\left(\hat{J}^P \hat{J}_P
-\psi^P\partial
\psi_P\right), \cr
G_{sl}=& {2\over k}\left(\psi^P\hat{J}_P -{i\over 3k}
\eps_{PQR}\psi^P
\psi^Q \psi^R\right).}}
Let us now turn to the $SU(3)/U(1)$ coset CFT.
We start from the $SU(3)$ affine superalgebra at level $k'=4k$ realized
as follows:
\eqn\opesu{\eqalign{K^A(z)K^B(w)\sim & {{k'\over 2}\delta^{AB} \over (z-w)^2}
+{if_{ABC} K^C(w) \over z-w}, \cr
K^A(z) \chi^B(w)\sim & {if_{ABC} \chi^C(w) \over z-w}, \cr
\chi^A(z)\chi^B(w)\sim & {{k'\over 2}\delta^{AB} \over  z-w}.}}
Here $A,B,C,D=1\dots 8$ and the structure constants $f_{ABC}$ are $f_{123}=1$,
$f_{147}=-f_{156}=f_{246}=f_{257}=f_{345}=-f_{367}={1\over 2}$ and
$f_{458}=f_{678}={\sqrt{3}\over 2}$. Since the metric is $\delta^{AB}$
we will not keep track of the upper or lower position of the
$SU(3)$ indices.
As before, we split the currents into their bosonic and fermionic parts:
\eqn\suk{K^A=\hat{K}^A-{i\over k'}f_{ABC}\chi^B \chi^C.}
The bosonic currents realize an affine algebra at level $k'-3$.

We now choose to mod out the $SU(3)$ by the $U(1)$ generated by $K^8$. The
$SU(2)$ subgroup generated by $K^1, K^2, K^3$ is orthogonal to this $U(1)$,
and thus survives as an affine algebra in the coset CFT.
The stress-energy tensor and the supercurrent of the coset CFT
are built as in \ks, using the decomposition $T_{SU(3)}=T_{SU(3)/U(1)}+
T_{U(1)}$, and similarly for the supercurrent $G$.
The stress-energy tensor reads:
\eqn\tcoset{\eqalign{T_{coset}=&{1\over k'}\left(\hat{K}^1\hat{K}^1+\dots
+\hat{K}^7\hat{K}^7\right) -{1\over k'}\left( \chi^1 \partial\chi^1 +\dots
+\chi^3 \partial \chi^3\right) \cr & -{1\over k'}\left(1-{3\over 2k'}\right)
\left( \chi^4 \partial\chi^4+\dots +\chi^7 \partial\chi^7\right)
+{2i\sqrt{3}\over {k'}^2} \hat{K}^8 \left( \chi^4 \chi^5+\chi^6\chi^7\right)
\cr & +{6\over {k'}^3} \chi^4\chi^5\chi^6\chi^7.}}

Our goal now is to build the spacetime supercharges. For that we would
like to construct spin-fields via bosonization following 
\ref\fms{D.~Friedan, E.~Martinec and S.~Shenker,
``Conformal Invariance, Supersymmetry And String Theory,''
Nucl.\ Phys.\  {\bf B271} (1986) 93.}.
Note that since we are dealing with a coset and not with a group manifold,
the fermions are generically not free. Of course since the $SU(2)$ is
preserved as an affine symmetry, the fermions belonging to it are free.
Despite the above remark, we proceed to bosonize the 10 fermions into
5 bosons. This will actually uncover the interesting structure of the
above coset model. Define:
\eqn\boso{\partial H_1 =  {2\over k} \psi^1 \psi^2, \quad
\partial H_2 =  {2\over k'} \chi^1 \chi^2, \quad
i \partial H_3 =  {1\over k} \psi^3 \chi^3, \quad
\partial H_4 = {2\over k'} \chi^4 \chi^5, \quad
\partial H_5 = {2\over k'} \chi^6 \chi^7.}
The scalars $H_I$ are all canonically normalized:
$H_I(z)H_J(w)\sim -\delta_{IJ}\log (z-w)$.
Conversely, the fermions are given by:
\eqn\ferm{\psi^1={\sqrt{k}\over 2}\left(e^{iH_1}+e^{-iH_1}\right), \qquad
\psi^2={i\sqrt{k}\over 2}\left(e^{iH_1}-e^{-iH_1}\right),}
and similarly for $H_2$, $H_4$ and $H_5$, while:
\eqn\fermspe{\psi^3={\sqrt{k}\over 2}\left(e^{iH_3}-e^{-iH_3}\right), \qquad
\chi^3={\sqrt{k'}\over 2}\left(e^{iH_3}+e^{-iH_3}\right),}
recalling that $H_3^\dagger=-H_3$ and $k'=4k$.

In terms of these scalars, the total stress-energy tensor is:
\eqn\ttot{\eqalign{T=T_{sl}+T_{coset}=& {1\over k}\left(\hat{J}^1\hat{J}^1 +
\hat{J}^2\hat{J}^2 - \hat{J}^3\hat{J}^3\right)
+{1\over k'}\left(\hat{K}^1\hat{K}^1+\dots+ \hat{K}^7\hat{K}^7\right) \cr &
-{1\over 2}\left( \partial H_1 \partial H_1 + \partial H_2 \partial H_2
+\partial H_3\partial H_3\right)
+{i\sqrt{3}\over k'}\hat{K}^8\left( \partial H_4 + \partial H_5 \right)
\cr & -{1\over 2}\left( 1-{3\over 2k'}\right)
\left(\partial H_4 \partial H_4 + \partial H_5\partial H_5\right)
+{3\over 2k'}\partial H_4 \partial H_5.}}
Obviously, the scalars $H_4$ and $H_5$ are not free in the coset CFT.
However, it is also easy to see that there is a linear combination of
them which is free. This is what will enable us to build the $N=3$
spacetime superalgebra.

We thus write:
\eqn\hminus{H_\pm={1\over \sqrt{2}} ( H_4 \pm H_5).}
Our final expression for $T$ is therefore:
\eqn\tfin{\eqalign{T= & {1\over k}\left(\hat{J}^1\hat{J}^1 +
\hat{J}^2\hat{J}^2 - \hat{J}^3\hat{J}^3\right)
+{1\over k'}\left(\hat{K}^1\hat{K}^1+\dots+ \hat{K}^7\hat{K}^7\right) \cr &
-{1\over 2}\left( \partial H_1 \partial H_1 + \partial H_2 \partial H_2
+\partial H_3\partial H_3 +\partial H_- \partial H_- \right) \cr &
-{1\over 2}\left( 1-{3\over k'}\right) \partial H_+ \partial H_+
+{i\sqrt{6}\over k'}\hat{K}^8 \partial H_+ .}}
We conclude that $H_-$ is the fourth free scalar, 
namely that $\partial H_-$ is a primary field of weight 1.

We now write the worldsheet $N=1$ supercurrent, which will be
used to enforce the BRST condition on the spin fields.
The supercurrent for the coset CFT reads:
\eqn\gcoset{G_{coset}={2\over k'}\left(\chi^{\bar{a}} \hat{K}^{\bar{a}}
-{i\over 3k'} f_{\bar{a}\bar{b}\bar{c}} \chi^{\bar{a}}\chi^{\bar{b}}
\chi^{\bar{c}} \right),}
where $\bar{a}$ are indices in the coset $G/H$.
Putting together \sltg\ and \gcoset, substituting the
structure constants of $SU(3)$ and taking into account the bosonization
in the term trilinear in the fermions, we get the expression for
$G_{tot}=G_{sl}+G_{coset}$:
\eqn\gtot{\eqalign{G_{tot}=& {2\over k} \left(\psi^1 \hat{J}^1 +\dots
-\psi^3 \hat{J}^3\right) +{2\over k'} \left(
\chi^1 \hat{K}^1 +\dots +\chi^7 \hat{K}^7\right) \cr &
+{i\over \sqrt{k}} \left\{ \partial H_1 \left( e^{iH_3}-e^{-iH_3}\right)
-\half \left( \partial H_2 +{1\over \sqrt{2}} \partial H_- \right)
\left( e^{iH_3}+e^{-iH_3}\right) \right\} \cr &
+{1\over 2 \sqrt{k} }\left(e^{iH_2 - i\sqrt{2}H_-}-e^{-iH_2+i\sqrt{2}H_-}
\right) .}}

Before going on to the BRST condition for the spin-fields, we write
for completeness the expressions for the $SU(2)$ currents, which remain
primary fields of weight 1 in the coset CFT and can also be considered
as the upper components of the fermions $\chi^1, \chi^2, \chi^3$.
Writing $K^\pm =K^1 \pm i K^2$ and similarly for the bosonic currents
and the fermions, we get:
\eqn\sucurr{\eqalign{K^\pm=& \hat{K}^\pm \mp e^{\mp i H_2}\left(
 e^{iH_3}+e^{-iH_3}\right) \pm e^{\mp i \sqrt{2}H_-} \cr
K^3 =& \hat{K}^3 -i \left(\partial H_2 +{1\over \sqrt{2}} \partial H_-
\right). }}
Note that since these currents are primaries of weight 1,
this could have been an alternative way of showing that $H_-$ is a free 
scalar.

\subsec{Physical operators and the spacetime algebra}

In order to construct the spacetime superconformal algebra we need,
in particular, to construct physical supercharges 
which we choose to write in the $-1/2$ picture \fms: 
\eqn\such{Q \propto \oint e^{-\varphi/2}u^{\alpha}S_\alpha (z) dz.}
Here $S_\alpha$ is a basis of spin-fields, $u^{\alpha}$ are constants, 
and $\varphi$ is the bosonized superconformal ghost.
The set of operators $e^{-\varphi/2}u^{\alpha}S_\alpha (z)$ 
should be BRST invariant and mutually local.
We choose a basis of spin-fields
\eqn\genspf{S_{[\eps_1 \eps_2 \eps_3 \eps_-]}=e^{{i\over 2}(\eps_1 H_1+
\eps_2 H_2 +\eps_3 H_3 +\eps_- \sqrt{2} H_-)},}
where $\epsilon_I=\pm 1$. Because $H_-$ is a free scalar, 
these 16 spin-fields are primaries of weight 5/8 and, therefore, 
$e^{-\varphi/2}u^{\alpha}S_\alpha (z)$ are primaries of weight 1,
as they should be.

The super BRST condition on $e^{-\varphi/2}u^{\alpha}S_\alpha$
further requires that there
will be no ${(z-w)^{-3/2}}$ singular terms in the OPE 
of $u^{\alpha}S_\alpha$ with the supercurrent
$G_{tot}$ (note that the only dangerous terms in $G_{tot}$ are the ones
trilinear in the fermions, i.e. the second and third lines in \gtot).
This leaves 8 combinations $u^{\alpha}S_\alpha$ 
out of the 16 spin-fields \genspf.
The GSO condition, i.e. mutual locality, further leads to one of two choices 
of chirality: $\eps_1 \eps_2 \eps_3 =-1$ or $\eps_1 \eps_2 \eps_3 =1$, 
under which 6 or 2 of the combinations 
$u^{\alpha}S_\alpha$ survive, respectively.

Explicitly, 
the outcome of the computation is the following. For spacetime chirality
$\eps_1 \eps_2 \eps_3 =-1$, we get 6 physical
spin-fields:
\eqn\spin{\eqalign{S^+_{\half}\ =&\ S_{[----]} \cr
S^-_{\half}\  =&\  S_{[-+++]} \cr
S^3_{\half}\ =&\ \half ( S_{[-++-]} -S_{[---+]}) \cr
S^+_{-\half}=&\ S_{[+-+-]} \cr
S^-_{-\half} =&\  S_{[++-+]} \cr
S^3_{-\half} =&\ \half ( S_{[++--]} -S_{[+-++]}).}}
The lower and upper labels of $S^a_r$ denote respectively the quantum
numbers of the global $SL(2)$ and $SU(2)$ symmetries, in the
$({\bf 2}, {\bf 3})$ representation, as can be checked by
taking the OPEs with the respective currents \decomp\ and \sucurr.
For the other spacetime chirality $\eps_1 \eps_2 \eps_3 =1$, 
we get 2 physical spin-fields:
\eqn\cospin{\eqalign{\tilde{S}_\half\ =& \ \half(S_{[--++]}+S_{[-+--]}) \cr
\tilde{S}_{-\half} =& \ \half(S_{[+++-]}+S_{[+--+]}).}}
It can be checked that the above spin-fields $\tilde{S}_r$
have regular OPEs with the $SU(2)$ currents.

We thus see that the choice of GSO projection will lead to different
amounts of supersymmetry in spacetime.
Namely, in a type II background, the projection in the left and right 
moving sectors of the worldsheet CFT determine, respectively, the amount of
supersymmetry in the left and right moving sectors
of the spacetime CFT. Specifically, the different GSO projections
would lead in type IIA to $N=(3,1)$ or
$N=(1,3)$, and in type IIB to $N=(3,3)$ or $N=(1,1)$.
In the heterotic string, the different GSO projections in the worldsheet
supersymmetric sector lead to $N=(3,0)$ or $N=(1,0)$ in spacetime.\foot{%
Note that these examples provide, in particular, a construction of $N=1$ 
supersymmetry in spacetime which is not 
a $Z_2$ orbifolding of the $N=2$ construction 
of \gr.
If it was such an orbifold, each of the supercharges would split
into two BRST invariant pieces, leading to a total of 4 physical spin-fields,
in contrast with the result \cospin.}

The generators of the spacetime global $N=3$ superconformal algebra
are the following:
\eqn\spops{\eqalign{L_{\pm 1}=&-\oint dz J^\pm (z), \qquad \qquad \qquad \quad
L_0 = - \oint dz J^3 (z) \cr
T_0^\pm = & \oint dz K^\pm (z), \qquad \qquad \quad \quad \quad \quad \!
T_0^3 = \oint dz K^3(z) \cr
Q_\half^\pm =&  \oint dz e^{-\varphi/2} S_\half^\pm(z), \quad \qquad \;
\qquad Q_\half^3 = \oint dz e^{-\varphi/2} S_\half^3 (z)\cr
Q_{-\half}^\pm =&   \oint dz e^{-\varphi/2} S_{-\half}^\pm(z), \qquad \;
\qquad Q_{-\half}^3 =  \oint dz e^{-\varphi/2} S_{-\half}^3(z) ,}}
where we omit the normalization and the cocycle factors in the
definition of the $Q$'s.
These operators close the global part of the $N=3$ superconformal algebra (up
to picture changing), in the NS sector:
\eqn\three{\eqalign{ [L_m, L_n] =& (m-n) L_{m+n} \cr
[T^a_0, T^b_0]=& i \eps^{abc} T^c_0 \cr
[L_m, T^a_0] = & 0 \cr
[L_m, Q_r^a ] = & \left(\half m - r\right) Q^a_{m+r} \cr
[T^a_0, Q_r^b] = &i \eps^{abc} Q_r^c \cr
\{ Q_r^a, Q_s^b\} =& 2 \delta^{ab} L_{r+s} +i \eps^{abc} (r-s) T^c_{r+s}, }}
where $m,n=0, \pm1$, $a,b,c=1,2,3$ and $r,s=\pm\half$.
Of course this model reproduces the full $N=3$ superconformal algebra.
The higher modes can be built as in \refs{\gks,\kuse}.
For instance, we can first construct all the $L_n$.
Then acting with them
on $T^a_0$ and $Q^a_{\pm\half}$ one gets all the $T^a_n$ and $Q^a_r$ higher
modes. To close the algebra an additional fermionic field is needed,
all the modes of which are obtained from commutators of  $T^a_n$ and $Q^a_r$.
The full algebra appears for example in \refs{\ss,\yis}.

For completeness, we also write the (global)
algebra for the other GSO projection,
that is an $N=1$ superconformal algebra together with an affine $SU(2)$
which acts trivially on the supercharges. The supersymmetry generators
are given in this case by:
\eqn\one{\tilde{Q}_{\pm \half} = \oint dz e^{-\varphi/2}
\tilde{S}_{\pm \half}(z), }
and the algebra is:
\eqn\onealg{\eqalign{ [L_m, L_n] =& (m-n) L_{m+n} \cr
[T^a_0, T^b_0]=& i \eps^{abc} T^c_0 \cr
[L_m, T^a_0] = & 0 \cr
[L_m, \tilde{Q}_r] = & \left(\half m - r\right) \tilde{Q}_{m+r} \cr
[T^a_0, \tilde{Q}_r] = & 0 \cr
\{ \tilde{Q}_r, \tilde{Q}_s\} =& 2 \delta^{ab} L_{r+s} . }}
Again, using the higher modes of $L_n$ one can generate the higher modes
of the other operators, together with the fermionic superpartners of
the affine currents (see \yis\ for an analogous construction).

Let us conclude this section by a brief comment on  a special case, 
when the level of the $SU(3)$ is $k'=3$
(the minimal level allowed by unitarity). 
We can decompose the coset CFT $[SU(3)/U(1)]$ into
the product $[SU(2)] \times [SU(3)/(SU(2)\times U(1))]$. The central
charge of the second piece is $c=0$ in the $k'=3$ case, thus
the whole model reduces to string propagation on $SL(2)_{3/4} \times
SU(2)_3$. If we consider the six dimensional model $SL(2)_k\times
SU(2)_{k'}$, criticality enforces ${1\over k} -{1\over k'}=1$. The
only combination which might allow $N=3$ in spacetime (i.e. which verifies
$k'=4k$) is the one above, and our analysis indeed shows that it has
$N=3$. We will comment more on this case later.

\vskip 1cm

\newsec{Superstring theory on $AdS_3\times SO(5)/SO(3)$ }
The construction of the $N=3$ superconformal algebra in this case
follows closely the steps of the previous section. We shall therefore
be more schematical, and focus on the specifics of this model.

The $SO(5)$ current algebra looks the same as \opesu, at the same
level $k'=4k$, but now the indices are $A,B,C,D=1\dots 9,0$ and the
structure constants $f_{ABC}$ are $f_{123}=f_{456}=1$,
$f_{170}=f_{189}=-f_{279}=f_{280}=f_{378}=f_{390}=\half$ and
$-f_{470}=f_{489}=-f_{579}=-f_{580}=f_{678}=-f_{690}=\half$.
We work in the basis where the two orthogonal $SO(3)$ subgroups
of $SO(5)$ are generated respectively by $K^1, K^2, K^3$ and
$K^4, K^5, K^6$. We will mod out by the second one, leaving the
first one as the R-symmetry.

As before, we straightforwardly bosonize the 10 fermions in the $SL(2)$ and
in the coset, to get:
\eqn\bososo{\partial H_1 =  {2\over k} \psi^1 \psi^2, \quad
\partial H_2 =  {2\over k'} \chi^1 \chi^2, \quad
i \partial H_3 =  {1\over k} \psi^3 \chi^3, \quad
\partial H_4 = {2\over k'} \chi^7 \chi^8, \quad
\partial H_5 = {2\over k'} \chi^9 \chi^0.}
We can now write the total stress-energy tensor in terms of them as
(we also use \hminus):
\eqn\ttotso{\eqalign{T=& {1\over k}\left(\hat{J}^1\hat{J}^1 +
\hat{J}^2\hat{J}^2 - \hat{J}^3\hat{J}^3\right)
+{1\over k'}\left(\hat{K}^1\hat{K}^1+\dots+ \hat{K}^3\hat{K}^3
+\hat{K}^7\hat{K}^7 +\dots + \hat{K}^0\hat{K}^0\right) \cr &
-{1\over 2}\left( \partial H_1 \partial H_1 + \partial H_2 \partial H_2
+\partial H_3\partial H_3 +\partial H_+\partial H_+ \right)
-{1\over 2}\left( 1-{3\over k'}\right) \partial H_- \partial H_- \cr &
+{i\sqrt{2}\over k'}\hat{K}^6 \partial H_-
-{1\over k'}\hat{K}^4 (e^{i\sqrt{2}H_-}-e^{-i\sqrt{2}H_-})
-{i\over k'}\hat{K}^5 (e^{i\sqrt{2}H_-}+e^{-i\sqrt{2}H_-})
.}}
Now $H_+$ is the free scalar.  

The analogous expression for the total supercurrent is:
\eqn\gtotso{\eqalign{G_{tot}=& {2\over k} \left(\psi^1 \hat{J}^1 +\dots
-\psi^3 \hat{J}^3\right) +{2\over k'} \left(
\chi^1 \hat{K}^1 +\dots +\chi^3 \hat{K}^3
+\chi^7 \hat{K}^7 +\dots +\chi^0 \hat{K}^0\right) \cr &
+{i\over \sqrt{k}} \left\{ \partial H_1 \left( e^{iH_3}-e^{-iH_3}\right)
-\half \left( \partial H_2 +{1\over \sqrt{2}} \partial H_+ \right)
\left( e^{iH_3}+e^{-iH_3}\right) \right\} \cr &
-{1\over 2 \sqrt{k} }\left(e^{iH_2 - i\sqrt{2}H_+}-e^{-iH_2+i\sqrt{2}H_+}
\right) .}}
The $SU(2)$ currents are:
\eqn\socurr{\eqalign{K^\pm=& \hat{K}^\pm \mp e^{\mp i H_2}\left(
 e^{iH_3}+e^{-iH_3}\right) \mp e^{\mp i \sqrt{2}H_+} \cr
K^3 =& \hat{K}^3 -i \left(\partial H_2 +{1\over \sqrt{2}} \partial H_+
\right). }}
The solutions to the BRST invariance conditions on the spin-fields
$S_{[\eps_1\eps_2\eps_3\eps_+]}$ are, for the $\eps_1\eps_2\eps_3=-1$
GSO projection:
\eqn\spinso{\eqalign{S^+_{\half}\ =&\ S_{[----]} \cr
S^-_{\half}\  =&\  S_{[-+++]} \cr
S^3_{\half}\ =&\ \half ( S_{[-++-]} +S_{[---+]}) \cr
S^+_{-\half}=&\ S_{[+-+-]} \cr
S^-_{-\half} =&\  S_{[++-+]} \cr
S^3_{-\half} =&\ \half ( S_{[++--]} + S_{[+-++]}).}}
For the other chirality, $\eps_1\eps_2\eps_3=1$, we get:
\eqn\cospinso{\eqalign{\tilde{S}_\half\ =& \ \half(S_{[--++]}-S_{[-+--]}) \cr
\tilde{S}_{-\half} =& \ \half(S_{[+++-]}-S_{[+--+]}).}}
{}From the above spin-fields and currents, the construction of the $N=3$
(or $N=1$ according to the GSO projection) algebra proceeds in exactly
the same manner as in the former case.

\newsec{$N=3$ superalgebra as an enhancement of $N=2$}
Since the $N=3$ superconformal algebra has the $N=2$ superalgebra as
a subalgebra, and since general conditions for the appearance of the
latter are known \gr, it is natural and instructive to investigate the
relation between the two constructions.

In \gr\ it was found that a general condition for having $N=2$
superconformal algebra in spacetime for a background of the form
$AdS_3 \times \NN$, is the existence of an affine $U(1)$ current
in $\NN$, such that $\NN / U(1)$ has $N=2$ worldsheet supersymmetry.
It was noted there that
the $U(1)$ must be chosen carefully in the cases where enhancement
to $N>2$ is expected, in order to embed the $N=2$ construction
into the explicit construction of the larger algebra.

We now proceed to show that in our two cases there is only one choice of the
complex structure in $\NN / U(1)$, where the $U(1)$ is the Cartan
subalgebra of the $SU(2)$, that leads to an $N=2$ construction
which is a subalgebra of the $N=3$ algebra constructed in the previous
sections.

The general construction of the $N=2$ superconformal algebra for
coset models \ks\ leads to the following $U(1)$ R-current:
\eqn\rcurr{J_R= {i\over k'} h_{\bar{a}\bar{b}}\chi^{\bar{a}}\chi^{\bar{b}}
+{1\over k'} h_{\bar{a}\bar{b}} f_{\bar{a}\bar{b}C}\left(\hat{K}^C
-{i\over k'} f_{C\bar{d}\bar{e}}\chi^{\bar{d}}\chi^{\bar{e}}\right).}
The index $C$ can run over both $H$ and $G/H$.
The complex structure $h_{\bar{a}\bar{b}}$ has to satisfy conditions
that can be found in \ks.

The construction of $N=2$ supercharges then proceeds as follows 
\gr.\foot{Note that this is not the
standard construction \ref\bd{T.~Banks, L.~J.~Dixon, D.~Friedan and 
E.~Martinec, ``Phenomenology And Conformal Field Theory 
Or Can String Theory Predict The Weak Mixing Angle?,''
Nucl.\ Phys.\  {\bf B299} (1988) 613; T.~Banks and L.~J.~Dixon,
``Constraints On String Vacua With Space-Time Supersymmetry,''
Nucl.\ Phys.\  {\bf B307} (1988) 93.}
with respect to $N=2$ supersymmetry on the worldsheet.}
We present a canonically normalized scalar $H_0$:
\eqn\hzero{i\sqrt{3}\partial H_0= J_R -{4\over k'}K^3, }
which is used to construct the spin-fields:
\eqn\spingr{S=e^{{i\over 2}(\eps_1 H_1 +\eps_3 H_3 +\eps_0 \sqrt{3}H_0)},}
where $H_1$ and $H_3$ are built as before \boso. The BRST condition will
pick up 4 of the above spin-fields as physical.%
\foot{Note that the BRST condition in this construction
leads to a definite GSO projection (i.e. no BRST invariant
spin-fields of the other chirality are explicitly constructed).}
This construction of
$N=2$ will be embedded in our $N=3$ constructions provided that
the above spin-fields \spingr\ can be rewritten as special cases of
\genspf.

Let us consider first the $SU(3)/U(1)$ case. Here we have to look for
complex structures of the six dimensional manifold $SU(3)/U(1)^2$.
We take the Cartan subalgebra of the $SU(3)$ to be generated by
$K^3$ and $K^8$. We find three possible complex structures.

The first complex structure is given by $h_{12}=h_{45}=h_{67}=1$, and leads
to an R-current of the form:
\eqn\rbada{J_R'=i\partial H_2 +i\sqrt{2}\partial H_+ +{2\over k'}K^3
+{2\sqrt{3}\over k'}K^8.}
Note that the above expression is such that both $\partial H_+$ and $K^8$
will appear in the definition \hzero\ of $H_0$, and therefore the spin-fields
\spingr\ cannot be matched to \genspf.

The second complex structure is given by $h_{12}=-h_{45}=-h_{67}=1$, and leads
to an R-current of the form:
\eqn\rbadb{J_R''=i\partial H_2 -i\sqrt{2}\partial H_+ +{2\over k'}K^3
-{2\sqrt{3}\over k'}K^8.}
The same remark as above applies here. 
Moreover these two currents are the sum
and difference of the $N=2$ $U(1)$ R-currents that one can get
by decomposing this coset CFT into $[SU(2)/U(1)]
\times [SU(3)/(SU(2)\times U(1))]$. We will explain shortly why this
direct-product decomposition cannot lead to the enhancement to $N=3$.

The third complex structure is given by $h_{12}=h_{45}=-h_{67}=1$, and leads
to an R-current of the form:
\eqn\rgood{J_R=i\partial H_2 +i\sqrt{2}\partial H_- +{4\over k'}K^3.}
The boson constructed as in \hzero\ now reads $\sqrt{3}H_0=H_2+\sqrt{2} H_-$.
The spin-fields \spingr\ are thus exactly of the form \genspf\ with
$\eps_2=\eps_-$. The BRST invariant ones are exactly the $N=2$
subalgebra generators of \spin, namely $S^\pm_{\pm {1\over 2}}$.

Moving to the $SO(5)/SO(3)$ case, we have to consider the complex
structure of $SO(5)/(SO(3)\times SO(2))$, where the $SO(2)$ 
is again the Cartan subalgebra 
of the remaining $SO(3)$, generated by $K^3$.
This case is different, as there is only one possible complex structure,
given by $h_{12}=h_{78}=h_{90}=1$. The associated R-current is:
\eqn\rgoodso{J_R=i\partial H_2 +i\sqrt{2}\partial H_+ +{4\over k'}K^3.}
As for \rgood, the BRST invariant spin-fields constructed 
according to \gr\ are
the $N=2$ supercharges $S^\pm_{\pm {1\over 2}}$ of \spinso.
The presence of only one complex structure in this case (as opposed
to three in the previous one) is due to the
fact that the four dimensional coset $SO(5)/SO(4)$ has no complex structure.

\newsec{General conditions for obtaining $N=3$}
The above discussion leads us to present general conditions for the
appearance of the $N=3$ superconformal algebra in the context of
string theory on $AdS_3 \times \NN$.
Such a background leads to $N=3$ superconformal algebra in spacetime
provided that:
\item{\it (i)} $\NN$ has an affine $SU(2)$ current algebra at level
$k'=4k$, where $k$ is the level of $SL(2)$.
\item{\it (ii)} $\NN/U(1)$ has $N=2$ worldsheet supersymmetry, where
$U(1)$ is the Cartan subalgebra of the above $SU(2)$. This condition
alone allows one to construct an $N=2$ superconformal algebra in
spacetime (for a definite GSO projection).
\item{\it (iii)} This spacetime
$N=2$ algebra is enhanced to $N=3$ if the scalar
$H_0$ constructed as in \hzero\ can be decomposed as
$\sqrt{3}H_0 =H_2 + \sqrt{2} \tilde{H}_0$, where $H_2$ derives from
the bosonization of the two remaining charged
fermions of the $SU(2)$, and
$\tilde{H}_0$ is orthogonal to it.

\noindent
Interestingly, these conditions imply as a by-product that for the 
opposite GSO projection we also get supersymmetry in spacetime, 
namely $N=1$.

Let us present the proof by constructing the $N=3$ superalgebra
generators given the above conditions.
Recall that besides the scalar \hzero, we also define \gr\ the scalars
$\partial H_1={2\over k}\psi^1 \psi^2$ and $i\partial H_3 ={1\over k}
\psi^3 \chi^3$. The existence of the affine $SU(2)$, of which $\chi^3$
is the lower component of the Cartan generator, allows us to define
also $\partial H_2= {2\over k'}\chi^1\chi^2$.
Consider now the currents $K^3$ and $K^\pm$.
Since they form an $SU(2)$ supersymmetric WZW model (embedded
inside the CFT on $\NN$), they can be split into orthogonal pieces:
\eqn\ususplit{K^3=\tilde{K}^3-i\partial H_2, \qquad \qquad
K^\pm=\tilde{K}^\pm \mp {2\over \sqrt{k'}}e^{\mp i H_2} \chi^3.}
We start now by noting that condition {\it (iii)} implies the following 
(making use of \hzero):
\eqn\htilde{i\sqrt{2}\partial \tilde{H}_0(z) K^3(w) \sim
-{1\over (z-w)^2}.}
This means that $K^3$ can be split further:
\eqn\split{K^3=\hat{K}^3-{i\over \sqrt{2}} \partial \tilde{H}_0
- i\partial H_2,}
where $\hat{K}^3$ has a regular OPE with $\tilde{H}_0$ (and of course
$H_2$).
Similarly, the currents $K^\pm$ also split into a `bosonic' part $\hat{K}^\pm$
which realizes an affine $SU(2)_{k'-3}$, an $SU(2)_1$ part built from
$\tilde{H}_0$ and the usual fermionic $SU(2)_2$ piece:
\eqn\splitpm{K^\pm=\hat{K}^\pm \mp e^{\mp i\sqrt{2}\tilde{H}_0}
\mp e^{\mp i H_2} (e^{iH_3} + e^{-iH_3}).}
We can now construct the 4 physical spin-fields as in \gr. Note that
the presence of the full $SU(2)$ is irrelevant in this step. Using the
spin-fields of the form \spingr, we get
$S_{[\eps_1 \eps_3 \eps_0]}=S_{[---]},\ S_{[-++]},\
S_{[++-]},\ S_{[+-+]}$. Splitting $\sqrt{3}H_0=H_2+\sqrt{2}\tilde{H}_0$,
we can rewrite them as:
\eqn\spina{S_{[\eps_1 \eps_2 \eps_3 \tilde{\eps}_0]}=
S_{[----]},\; S_{[-+++]},\; S_{[+-+-]},\; S_{[++-+]}.}
We now find the additional two BRST invariant supercharges, by acting on
the above spin-fields with the $SU(2)$ ladder operators $K^\pm$, which
are also BRST invariant (as upper components of primaries of weight $1/2$).
The result is:
\eqn\spinb{\half(S_{[---+]}+S_{[-++-]}),\qquad \half(S_{[+-++]}+S_{[++--]}).}
The set of spin-fields \spina-\spinb\ matches exactly the ones found in
the cases detailed in the previous sections (note that the apparent sign
difference with \spin\ and \sucurr\ can be absorbed in a redefinition of
fields; in section  3
we preferred to stick to the usual Gell-Mann basis of $SU(3)$).

Defining the spacetime operators as in \spops, one can show that the
$N=3$ superconformal algebra closes.

The above proof builds upon the existence of the $N=2$ superalgebra,
enhancing it to $N=3$ using the $SU(2)$ currents.
An alternative way of building the $N=3$ superalgebra, which also 
reveals the existence of the $N=1$ superalgebra for the other GSO
projection, is to decompose the supercurrent of the CFT on $\NN$
into an $SU(2)$ part and a $\NN/SU(2)$ one.
It can then be used to directly find all of the 8 physical spin-fields, 6 of
one chirality and 2 of the other.
The $SU(2)$ part of the supercurrent is:
\eqn\gsutwo{G_{SU(2)}={2\over k'}\left(\half \chi^+ \tilde{K}^- +\half \chi^-
\tilde{K}^+ + \chi^3 \tilde{K}^3 -{2i\over k'}\chi^1 \chi^2 \chi^3\right).}
Using \ususplit, \split, \splitpm, and the bosonization, the relevant part
of $G_{SU(2)}$ for the BRST condition (i.e. the one that might lead
to $(z-w)^{-3/2}$ singular terms in the OPE with the spin-fields) is:
\eqn\relev{G_{SU(2)}=\dots + {1\over \sqrt{k'}}\left\{-i\left(\partial H_2
+{1\over \sqrt{2}} \partial \tilde{H}_0\right)(e^{iH_3} + e^{-iH_3})
-(e^{iH_2-i\sqrt{2}\tilde{H}_0} - e^{-iH_2+i\sqrt{2}\tilde{H}_0})\right\}.}
The first piece will give rise to a $(z-w)^{-3/2}$ singularity only when
$\eps_2=\tilde{\eps}_0$, while 
the second piece will do so only when
$\eps_2=-\tilde{\eps}_0$. Choosing $\eps_2=\tilde{\eps}_0$, we get
4 physical spin-fields of the same chirality. For $\eps_2=-\tilde{\eps}_0$
we get 4 physical spin-fields, two of each chirality.

It is worth noting that the direct product
$\NN=SU(2)_{k'}\times \NN'$,
which manifestly fulfills conditions {\it (i)} and {\it (ii)},
does not fulfill the third condition (except for one case which will
be discussed shortly). The reason for this is the following.
If $H_0$ fulfills condition {\it (iii)}, it is straightforward to compute
the OPE of $i\partial H_2$ with $J_R$, the R-current of $\NN/U(1)$,
the result being:
\eqn\condjr{i\partial H_2(z) J_R(w) \sim {1-{4\over k'} \over (z-w)^2}.}
However, if $\NN=SU(2)_{4k}\times \NN'$, then:
\eqn\jrsum{J_R = J_R^{SU(2)/U(1)} +J_R'=i\partial H_2 +{2\over k'}K^3
+J_R'.}
Since $i\partial H_2$ has a regular OPE with $J_R'$, its OPE
with $J_R$ in \jrsum\ gives a double
pole with a residue of $\left(1-{2\over k'}\right)$ instead of \condjr.
Thus such a CFT with a direct product $SU(2)$ factor will
not lead to $N=3$ in spacetime through the mechanism described above.

This seems to contradict what was noted about the limiting $k'=3$ case,
which was reduced to $SL(2)_{3/4}\times SU(2)_3$. This is resolved by
noting that we can always take $J_R$ to $-J_R$ (this amounts to changing
an overall sign in the complex structure, see \rcurr). Doing this,
the OPE of $i\partial H_2$ with $J_R$ gives a residue of
$-\left(1-{2\over k'}\right)$, which is equal to $\left(1-{4\over k'}\right)$
only if $k'=3$, that is in this particular case.
For completeness, we sketch the construction in this case. Using \hzero\
and \jrsum,
we write:
\eqn\sutwo{i\sqrt{3}\partial H_0=-J_R -{4\over 3}K^3=-i\partial H_2 -2K^3=
i\partial H_2 -2\hat{K}^3.}
Now we use the fact that the
bosonic $SU(2)$ WZW model at level 1 can be reformulated as
the CFT of a free scalar at its self-dual radius. Denoting this scalar
by $\tilde{H}_0$, we have $\hat{K}^3=-{i\over \sqrt{2}}\partial \tilde{H}_0$
and $\hat{K}^\pm=e^{\mp i \sqrt{2} \tilde{H}_0}$. It is thus clear
that $H_0$ fulfills condition {\it (iii)}, and the construction of
the $N=3$ superalgebra then proceeds as before.

\newsec{Comments}
Let us first comment on the relation between the general construction
of the previous section and the two specific coset CFTs discussed
before. It is possible
to check that the only 7 dimensional cosets $\NN$ which
have an $SU(2)$ symmetry (and do not factorize into a direct product
$SU(2)\times \NN'$)  are precisely $\NN_1$ and $\NN_2$ of eqs.~\su\
and \so. 
It is interesting that condition {\it (i)} of the previous
section together with the requirement of having a semiclassical 
$k \rightarrow \infty$ 
limit automatically lead to models which fulfill the two remaining 
conditions. Considering 7 dimensional group manifolds, only the case
$\NN=SU(2)_{2k}\times SU(2)_{2k}\times U(1)$ satisfies condition {\it (i)},
where the $SU(2)_{4k}$ is the diagonal one. It is
straightforward to show that this manifold also satisfies conditions
{\it (ii)} and  {\it (iii)}.
This manifold actually
possesses large $N=4$ superconformal symmetry \efgt, 
of which $N=3$ is a subalgebra.
This fact was used in \yis\ to break $N=4$ to $N=3$ through 
a $Z_2$ orbifold construction.
It would be interesting to find the relation between this small set
of models which share the same spacetime superconformal symmetry. 

We conclude by commenting on the geometrical interpretation. It would
be nice to translate the conditions we impose on the CFT on $\NN$
into conditions on the geometry of the manifold. It is clear
as was commented before that the $S^3 \cong SU(2)$ has to be non-trivially
fibered over the 4 dimensional base $\NN/SU(2)$.
A related, but different, problem
was actually discussed in the literature \refs{\cast,\figu},
where conditions on Einstein 7-manifolds $\NN$
are found in order to get different amounts of supersymmetry when considering
11 dimensional supergravity on $AdS_4 \times \NN$. The condition
for getting $N=3$ in $AdS_4$
is that $\NN$ has a tri-sasakian structure
(in other words, the cone over it $C(\NN)$ is hyperK\"ahler; see for instance
\figu\ for the notions introduced hereafter). 
The above geometries are considered as near horizon geometries of M2-branes
at the singularity of the Ricci-flat cone $C(\NN)$ over such manifolds.
The tri-sasakian structure implies the presence of 3 Killing vectors forming an
$SO(3)$ algebra which rotates the 3 Killing spinors.
It turns out that the only 7-dimensional tri-sasakian manifolds (satisfying
some additional regularity conditions, see also 
\ref\fk{T.~Friedrich and I.~Kath,
``Seven-Dimensional Compact Riemannian Manifolds With Killing Spinors,''
Commun.\ Math.\ Phys.\  {\bf 133} (1990) 543.})
are exactly the cosets\foot{The case $S^3 \times
S^3 \times S^1$, which has $N=3$ as a subgroup
of large $N=4$, does not appear in the classification above since
its metric cannot be rescaled to become an Einstein manifold
because of the flat $S^1$ factor.} $SU(3)/U(1)$ and $SO(5)/SO(3)\cong
S^7$. In the second case, the quotient is taken as in section 4 
\ref\crw{L.~Castellani, L.~J.~Romans and N.~P.~Warner,
``Symmetries Of Coset Spaces And Kaluza-Klein Supergravity,''
Annals Phys.\  {\bf 157} (1984) 394; ``A Classification Of 
Compactifying Solutions For D = 11 Supergravity,''
Nucl.\ Phys.\  {\bf B241} (1984) 429.}. Note also
that $S^7$ is
trivially tri-sasakian since it has $N=8$ Killing spinors.

Tri-sasakian manifolds can be seen as $S^3$ fibrations over a base, 
in the above cases ${\bf CP}^2$ and $S^4$ respectively.
It is interesting to note that `squashing'
a tri-sasakian manifold (i.e. rescaling the fiber with respect
to the base) leads, for a definite value of the squashing parameter,
to another Einstein manifold having one Killing spinor (instead of 3) and
an unbroken $SO(3)$ algebra of Killing vectors which acts trivially
on the spinor.
This is reminiscent of our results, where the
different amounts of supersymmetry, $N=3$ or $N=1$, depend however on the GSO
projection. 

We should nevertheless stress that in spite of these similarities, there
are a few differences. For instance,
superstring theory
on $AdS_3 \times \NN$ does not require $\NN$ to be an Einstein 
manifold.\foot{%
This might explain why in the $SO(5)/SO(3)$ case discussed in section 4
we do not find $N=8$ supersymmetry but only $N=3$.}
Recall that the metric of the coset CFT sigma model, which can be obtained
by gauging the WZW model on the group $G$ and integrating out the gauge
fields, is not the same as the metric on the homogeneous $G/H$ coset space.
Thus presumably the direct relation between the two issues is more algebraic
in nature than geometrical.

\nref\ct{P.~M.~Cowdall and P.~K.~Townsend,
``Gauged supergravity vacua from intersecting branes,''
Phys.\ Lett.\  {\bf B429} (1998) 281,
hep-th/9801165.}
\nref\bps{H.~J.~Boonstra, B.~Peeters and K.~Skenderis,
``Brane intersections, anti-de Sitter 
spacetimes and dual superconformal  theories,''
Nucl.\ Phys.\  {\bf B533} (1998) 127,
hep-th/9803231.}
\nref\ggpt{J.~P.~Gauntlett, G.~W.~Gibbons, G.~Papadopoulos and P.~K.~Townsend,
``Hyper-K\"ahler manifolds and multiply intersecting branes,''
Nucl.\ Phys.\  {\bf B500} (1997) 133,
hep-th/9702202.}
\nref\pt{G.~Papadopoulos and A.~Teschendorff,
``Grassmannians, calibrations and five-brane intersections,''
hep-th/9811034.}
\nref\papa{G.~Papadopoulos,
``Rotating rotated branes,''
JHEP {\bf 9904} (1999) 014,
hep-th/9902166.}
Another question regards the brane configuration which might lead to
the models considered here in the near horizon limit.
Since we are dealing with pure NSNS backgrounds in type II theories, 
we expect such a brane configuration to involve fundamental strings
and NS5-branes intersecting on the string, and possibly at non-trivial angles.
Indeed it is known that the $AdS_3\times S^3\times T^4$ and the
$AdS_3\times S^3\times S^3\times S^1$ backgrounds are the near horizon
geometries of configurations involving respectively 
a fundamental string within a NS5-brane
\gks, and an additional NS5-brane intersecting the other orthogonally 
on the string \refs{\ct,\bps}.
The latter brane configuration can be generalized by introducing a non-trivial
angle between the two NS5-branes, 
but still requiring that some supersymmetries
are preserved. This problem has been studied from the supergravity solution
point of view in \refs{\ggpt,\pt}, and it can be reduced to
a problem of classifying the holonomy of an 8-dimensional manifold (the
manifold which is orthogonal to the string intersection). It turns out \ggpt\
that solutions preserving a fraction of 3/16 of the supersymmetries
are associated with 8-dimensional manifolds of holonomy 
$Sp(4)\cong U(2,{\bf H})
\cong SO(5) \subset SO(8)$, which are thus hyperK\"ahler. This configuration
would lead to $N=(3,3)$ supersymmetry in type IIA.\foot{The supersymmetry of
the full brane configuration in type IIB is further reduced. For instance,
the orthogonal intersection already preserves 
only 1/8 of the supersymmetries.}
This is still different from what we are looking for.
In \pt\ more general NS5-brane configurations are considered, which are
related to hyperK\"ahler manifolds with torsion. The torsion allows for
the holonomy to be different for the two chiralities of the spinors (again
in type IIA). In this set up, solutions which have $N=(3,1)$ supersymmetry
are found, the associated manifold having $Sp(4)$ holonomy for one
chirality and $Spin(7)$ holonomy for the other. It should be noted that
the latter solutions, as presented in \pt, do not include fundamental strings. 
The near horizon geometry of some of the configurations above is considered
in \papa, where it is found to be actually $AdS_3\times S^3\times S^3
\times S^1$ like in the orthogonal case (hinting towards a large $N=4$
dual CFT, instead of an $N=3$ one).

It is not straightforward to see how our coset manifolds could arise
as near horizon geometries, at least in this context. 
In order to investigate this problem, one might need to further characterize
the 8-dimensional hyperK\"ahler manifolds involved
in the brane configurations. Note that the latter are asymptotically flat,
and thus generically not in the
same class as the conical ones discussed previously, which were
related to the classification of tri-sasakian (coset) manifolds.

\nref\gkp{A.~Giveon, D.~Kutasov and O.~Pelc,
``Holography for non-critical superstrings,''
JHEP {\bf 9910} (1999) 035,
hep-th/9907178.}
An alternative to configurations with branes at angles, is to consider
the near horizon limit of NS5-branes wrapping on 4-cycles,
together with fundamental strings stretched along the unwrapped direction.
These are expected \gkp\ to be equivalent to a superstring on 
$AdS_3\times \NN$ where $\NN$ has an affine $U(1)$ symmetry
and $\NN/U(1)$ is an $N=2$ SCFT related to the geometry 
of the 4-cycle. For particular geometries, the ${\cal N}/U(1)$ SCFT
was identified with the infrared limit of $N=2$ Landau-Ginzburg (LG) 
models \gkp. 
In the examples considered in this work ${\cal N}/U(1)$
is $SU(3)/U(1)^2$ and $SO(5)/(SO(3)\times U(1))$.
Generically, these $N=2$ quotients do not have a LG description
(except for the lowest levels of the $SU(3)/U(1)^2$ case).
Therefore, the relation to brane configurations along these lines 
requires the understanding of the duality of \gkp\ in models of the form
$AdS_3 \times G/H$, where $G/(H\times U(1))$ is a generic $N=2$ quotient.

\bigskip
\noindent{\bf Acknowledgements:}
We are happy to thank 
S.~Elitzur, O.~Feinerman, D.~Kutasov, W.~Lerche, E.~Rabinovici, 
M.~Ro\v cek and A.~Tomasiello for discussions and helpful remarks.
A. G. also thanks the ITP in Santa Barbara for hospitality during
the initial stages of this work.
This work is supported in part by the Israel
Academy of Sciences and Humanities -- Centers of Excellence Program,
and by the BSF -- American-Israel Bi-National
Science Foundation.

\listrefs

\end